\documentclass[aps, prl,amsmath,amssymb,twocolumn,floatfix]{revtex4}

\newif\ifpdf\ifx\pdfoutput\undefined\pdffalse\else\pdfoutput=1\pdftrue\fi

\usepackage{graphicx}
\usepackage{bm}
\usepackage{epsfig}

\newcommand{\p}{^{0}}

\newcommand{\rv}{\bm{r}}

\begin{document}
\title{\bf Crystalline phases of polydisperse spheres}

\author{Peter Sollich}
\affiliation{King's College London, Department of Mathematics, Strand,
London WC2R 2LS, United Kingdom.}

\author{Nigel B. Wilding}
\affiliation{Department of Physics, University of Bath, Bath BA2 7AY, United Kingdom.}

%\date{\today}

\begin{abstract}

We use specialized Monte Carlo simulation methods and moment free energy
calculations to provide conclusive evidence that dense polydisperse
spheres at equilibrium demix into coexisting fcc phases, with more phases appearing as
the spread of diameters increases. We manage to track up to four
coexisting phases. Each of these is fractionated: it contains a narrower
distribution of particle sizes than is present in the system overall. We
also demonstrate that, surprisingly, demixing transitions can be nearly
continuous, accompanied by fluctuations in local particle size
correlated over many lattice spacings.

\end{abstract}

\maketitle
\setcounter{totalnumber}{10}

Suspensions of spherical colloids have long served as an experimentally 
accessible testing ground for our understanding of the liquid,
crystalline and glassy states of matter \cite{PUSEY,PUSEY09}. Such work 
is complemented by theory and simulation, which attempt to reproduce,
rationalize and predict experimental results. In so doing, it is common
to treat the suspension as an assembly of {\em identical} spheres.
But this neglects a key feature, namely that the chemical processes
by which real colloids are synthesized invariably produce particles
that have a spread of diameters, i.e.\ they are `polydisperse'. As is
becoming increasingly clear, polydispersity gives rise to a rich variety
of novel phenomena not observed in monodisperse systems
\cite{Sollich2002}. However, despite sustained attention, basic
questions remain concerning its effects on one of the most fundamental
aspects of any thermal system, namely the equilibrium phase behaviour. 

A case in point is the character of the thermodynamically stable
structures of size-disperse spheres in the dense regime, above typical
fluid densities. Polydispersity should act to destabilize a crystal
because of the difficulty of accommodating a range of particle sizes
within a single lattice structure; but there has been no definite answer
as to what stable structures arise instead.  Indeed, the nature and
extent of the influence of polydispersity both on the crystalline phases
and the location of the freezing line is controversial. On the
theoretical front, there is a diverse range of predictions of novel
phenomena including reentrant melting \cite{Bartlett1999}, an
`equilibrium glass' phase \cite{Chaudhuri2005}, and solid-solid
coexistence \cite{Bartlett1998,Sear1998,Yang2009}. Additionally, recent
simulation work has reported the occurrence of a partly crystalline
`inhomogeneous phase' within an approximate phase diagram based only on
equality of single-phase free
energies~\cite{Fernandez2007a,Fernandez2009}. Other simulations suggest
that the fluid-solid coexistence region terminates in a critical point
beyond which a disordered solid occurs \cite{NOGAWA2009}. On the
experimental side, studies of  colloidal systems observe that beyond a
certain `terminal' polydispersity no crystallization occurs on
experimental timescales \cite{PUSEY}, although it remains unclear
whether this is a true equilibrium effect or a manifestation of dynamic
arrest.

A crucial distinction between monodisperse and polydisperse systems at phase
coexistence is the ability of the latter to {\em fractionate} so that
the distribution of the particle diameters, $\sigma$, varies from phase
to phase \cite{fredrickson1998,Sollich2001,Erne2005}. If for a certain
phase (labeled $\alpha$), one counts the number density of particles having
diameters in the range $\sigma\ldots\sigma+d\sigma$, this serves to
define a density distribution $\rho^{(\alpha)}(\sigma)$.
Experimentally, however, for most complex fluids one has the constraint
that the overall distribution of sizes (across all phases) has a form
fixed by the synthesis of the fluid. This gives rise to a generalized
lever rule:
$\rho^{(0)}(\sigma)=\sum_\alpha\lambda_\alpha\rho^{(\alpha)}(\sigma)$,
with $\lambda_\alpha$ the fractional volume occupied by phase $\alpha$,
$\rho^{(0)}(\sigma)$ the `parent' density distribution and
$\rho^{(\alpha)}(\sigma)$ the `daughter' distributions. Since the form
of the parent is fixed, only its scale is free to vary, e.g.\ by
dilution with solvent, and one writes $\rho^{(0)}(\sigma)=n\p
f(\sigma)$, where $n\p$ is the total number density and $f(\sigma)$ is a
prescribed normalized shape function.  The polydispersity, $\delta$, is
then defined as the standard deviation of the parent distribution, in
units of its mean. 

% REVISION  Removed the stuff on definition of cloud and shadow curves

The diversity of theoretical and simulation findings stems from the
sensitivity of the results to the accuracy with which fractionation is
treated. Previous work has either disregarded fractionation entirely, or
used drastic (and differing) approximations to describe it. An exception
are moment free energy (MFE) theory calculations, which do account fully
for fractionation, and which have previously been reported by one of us
for hard spheres \cite{Fasolo2004}. These predict that increasing
polydispersity shifts the fluid-solid coexistence region to higher
number densities, but that neither reentrant melting nor a terminal
polydispersity occurs. Instead, the fluid can always split off a small
volume of dense phase whose size distribution is sufficiently narrow for
crystallization. Moreover, as one increases $n\p$ or $\delta$ within the
solid region, a succession of phase transitions is predicted in which
the system demixes into an ever greater number of
differently-fractionated  `daughter' phases. However, the MFE
calculation uses approximate free energy expressions, which for solids
are derived from those of binary mixtures and implicitly already assume
that all solids are fcc. Independent confirmation of its predictions is
then highly desirable, but has hitherto been lacking. In this Letter we
provide a definite answer to the question of the nature of the
equilibrium phase behaviour via state-of-the-art Monte Carlo (MC)
simulations, and compare with MFE calculations;  both fully provide for
fractionation and employ a fixed parent size distribution. 

%REVISION reworded to emphasise that SGCE solves the local sampling problem
% I've also stuck the cloud and shadow stuff in here.

In simulations the appropriate framework for observing genuine
equilibrium behaviour in dense polydisperse particles is the isobaric
semi-grand canonical ensemble \cite{Kofke1999,FrenkelSmit2002}. This
is the analog of a monodisperse $(N,p,T)$ ensemble where the
prevalence of different particle sizes is controlled by imposing in addition
chemical potential differences $\Delta\mu(\sigma)$ that are measured
relative to the chemical potential of some reference particle size. Monte Carlo
sampling of this ensemble
can exploit particle resizing moves to allow local sampling of the
size distribution without 
the need for particle diffusion (thus catering for fractionation
effects), while volume updates facilitate density fluctuations so that
the system can transform between phases. Our study is the first to
deploy this ensemble in the crystalline regime together with a method
for imposing a fixed overall parent distribution. This allows determination
of physically realistic phase behaviour including the boundaries of the
onset of coexistence (known as cloud curves) and daughter distributions.
Additionally we can calculate -- but do not show here -- shadow curves
which record the density and 
volume fraction of the new phase when coexistence first occurs. Cloud and shadow
curves do not coincide, demonstrating further the presence of
fractionation: new phases that appear generically have 
size distributions different from the parent \cite{Sollich2002}.
We combine 
the above techniques with the specialized phase switch Monte Carlo (PSMC)
method \cite{Wilding2000,Wilding2009a} for obtaining fluid-solid
coexistence properties. In both cases, the chemical 
potential differences $\Delta\mu(\sigma)$ are determined iteratively
to match the 
ensemble-averaged density distribution $\langle\rho(\sigma)\rangle$ to
the prescribed parent $\rho^{(0)}(\sigma)=n\p f(\sigma)$. At
coexistence, this is supplemented by an equal peak weight criterion for
the order parameter distribution to ensure that finite-size effects are
exponentially small  in system size \cite{buzzacchi2006,Wilding2009}. 

%REVISION New paragraph

We stress that the choice of ensemble and use of sophisticated sampling
and analysis techniques are crucial to observing qualitatively correct
phase behaviour in polydisperse systems. Use of standard
canonical \cite{Fernandez2007a} or microcanonical ensembles
\cite{Fernandez2009,NOGAWA2009} are unequal to the task and almost
certainly yield major artifacts. The reasons for this are three fold:
(i) the dynamics is too slow to allow fractionation on simulation
timescales; (ii) the sizes of the particles are fixed, which for a finite
system prevents daughter distributions assuming an arbitrary form as
they can in the thermodynamic limit; (iii) these ensembles necessarily
form interfaces between coexisting phases and for accessible particle
numbers one cannot hope to see multiple coexisting crystalline phases
when this occurs.

Our simulations consider a system of $256$ particles interacting via a
strongly repulsive pair potential
\begin{equation}
v(r_{ij})=\epsilon(\sigma_{ij}/r_{ij})^{12}\:,
\label{eq:softspheres}
\end{equation}
with particle distances $r_{ij}=|\rv_i-\rv_j|$ and
interaction radii $\sigma_{ij}=(\sigma_i+\sigma_j)/2$. The choice of
this potential rather 
than infinitely repulsive (hard) spheres is made on pragmatic grounds:
an MC contraction of the simulation box that leads to an infinitesimal
overlap of two hard spheres will always be rejected, so (particularly at
high densities) we can expect higher MC acceptance rates using this
`softer' potential. In common with hard spheres, the
monodisperse version of our model freezes into an fcc crystalline
structure \cite{Hoover1970,Wilding2009a}, and temperature only plays
the role of a scale: the thermodynamic state depends
not on $n\p$ and $T$ separately but only on the combination
$n\p(\epsilon/k_BT)^{1/4}$. Phase diagrams for different $T$ then
scale exactly onto one another, and we can fix
$\epsilon/k_BT=1$.

We consider parent size
distributions of the top-hat form:
\begin{equation}
f(\sigma)=\left\{
\begin{array}{ll}
(2c)^{-1} & \mbox { if $1-c\le \sigma \le 1+c$} \\
~~0      &  \mbox { otherwise }
\end{array}
\right. .
\label{eq:th}
\end{equation}
The width parameter $c$ controls the
polydispersity $\delta=c/\sqrt{3}$, and we have set the mean particle
diameter to 1. With these choices, and the interaction potential
(\ref{eq:softspheres}), our results are directly comparable to the
phase diagram of Ref.~\cite{Fernandez2007a} where neither
fractionation nor, at a more basic level, the presence of coexistence
regions of finite width was allowed for.

\begin{figure}[h]
\includegraphics[type=pdf,ext=.pdf,read=.pdf,width=0.95\columnwidth,clip=true]{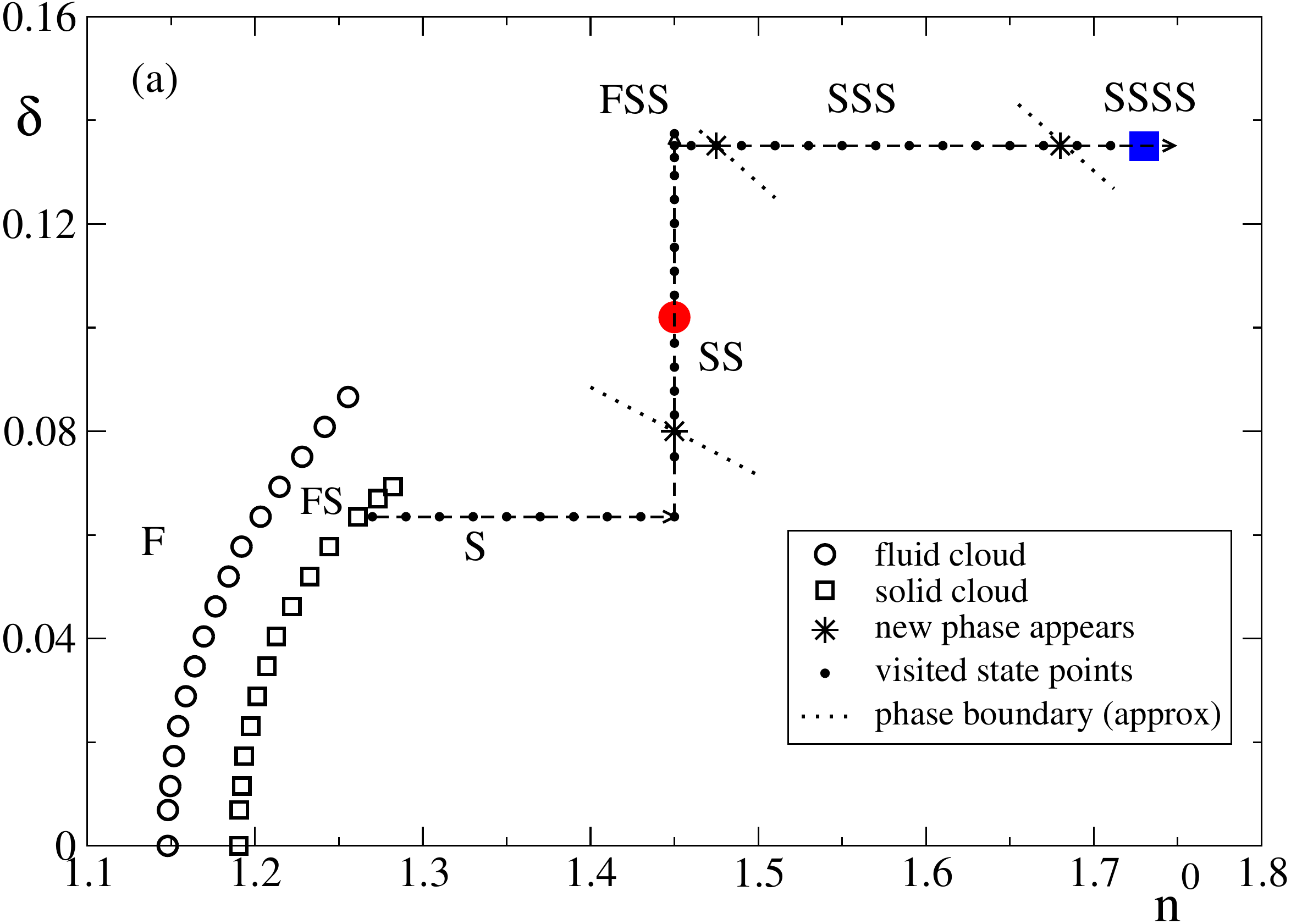}\\
\includegraphics[type=pdf,ext=.pdf,read=.pdf,width=0.95\columnwidth,clip=true]{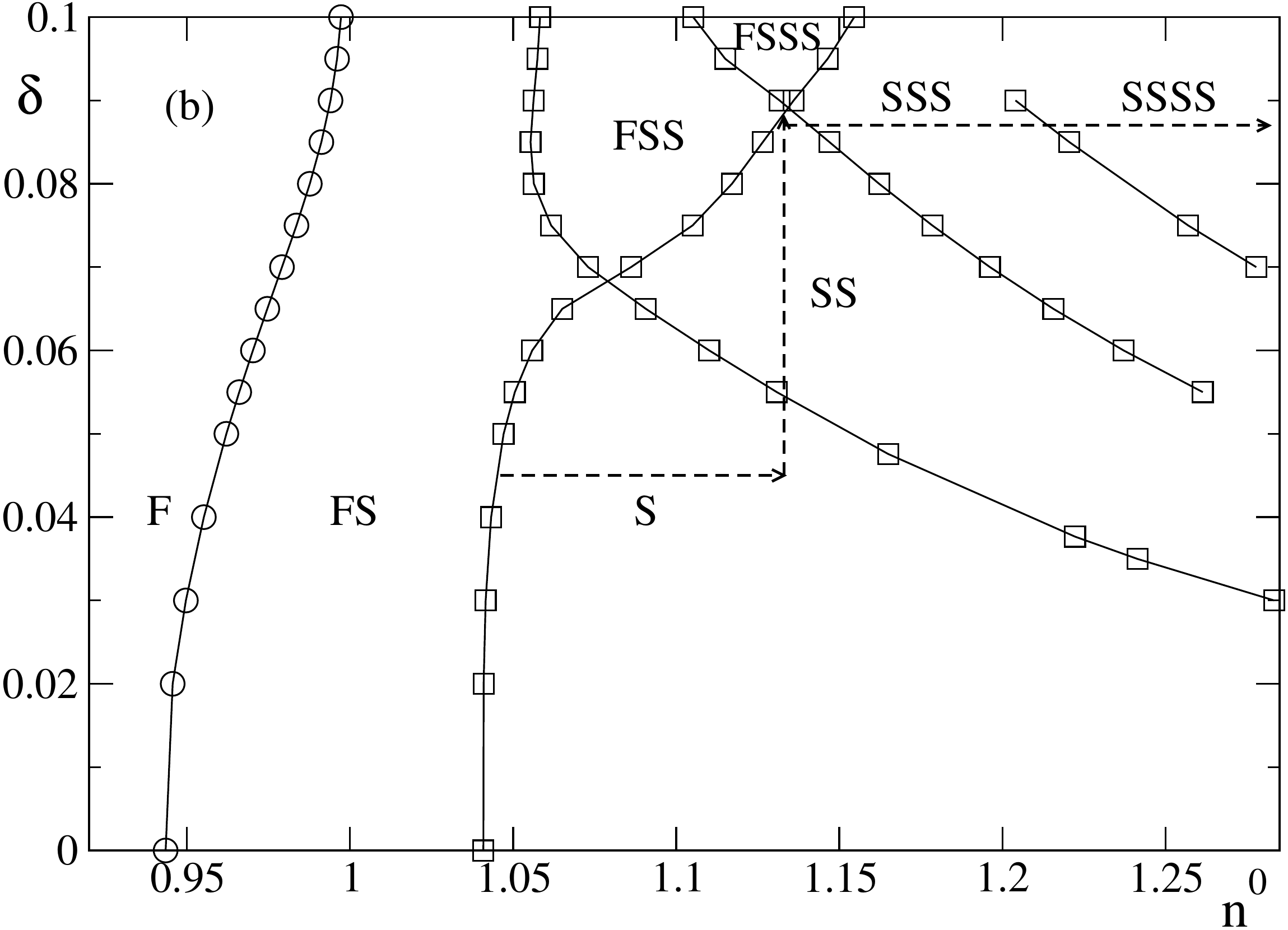}
%REVISION refers to colored points now included in figure
\caption{(Color online). {\bf (a)} Simulation results for the partial
phase diagram of the
model~(\ref{eq:softspheres}) with parent distribution~(\ref{eq:th}).
Asterisks: points where new solid phases appear; dashed lines: phase
boundary slopes found by histogram reweighting. F=fluid, S=solid.  
Colored symbols: state points considered in
Fig.~\protect\ref{fig:dendistsall}.
{\bf (b)} MFE calculation of phase diagram of hard spheres with the same
parent form. The dashed line shows a trajectory comparable to that
followed by the simulations.}
\label{fig:part_pd}
\end{figure}

%REVISION, dropped reference to shadow curves.

Using PSMC we have mapped the cloud curves of the fluid-solid
transition, using $\delta$ as our control parameter, up to
polydispersities of $\delta\approx 8.7\%$ on the fluid side and
$\delta\approx 7\%$ for the solid, both of which are in the typical
range for colloidal systems, but not so great that we expect to see 
exotic phases such as $AB_{13}$. As shown in Fig.~\ref{fig:part_pd}a, both
the fluid (circles) and the solid (squares) phase cloud densities shift
to higher $n\p$ as $\delta$ is increased, but without the sharp
narrowing that would be required for a reentrant melting scenario
\cite{Bartlett1999}.

\begin{figure}[h]
\vspace*{-2mm}
\includegraphics[type=pdf,ext=.pdf,read=.pdf,width=0.95\columnwidth,clip=true]{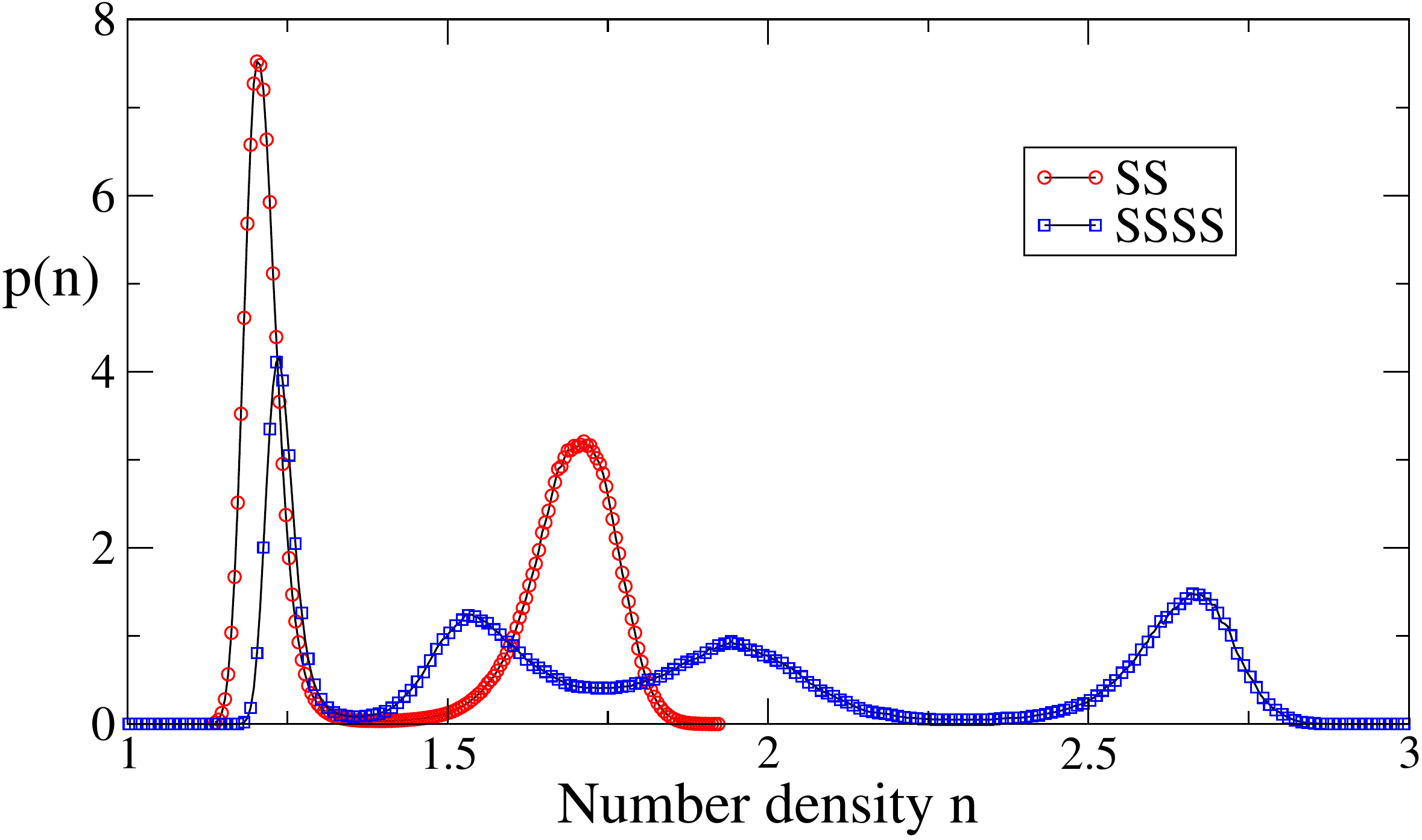}\\
\vspace*{-4mm}
%REVISION shortened by removing numerical values for state points and
% refering to points in fig 1. datasets color coded to match fig 1.
\caption{(Color Online). Distribution of the overall number density
$p(n)$ at the SS and SSSS statepoints indicated
by the colored symbols in Fig.~\protect\ref{fig:part_pd}a.
}

\label{fig:dendistsall}
\end{figure}

Turning now to the solid region, a comprehensive exploration of the
($n\p$-$\delta$) plane is impractical because of the relatively high 
computational cost of our specialized simulation technique. But we can
understand important qualitative features by following the dashed
trajectory included in Fig.~\ref{fig:part_pd}a. Along this path, we
monitored the state of the system via the probability distribution of
the fluctuating total number density $p(n)$, which serves as an order
parameter for phase changes. Starting from the fcc solid cloud point at
$\delta=6.3\%$, we initially increased $n\p$ in a stepwise fashion
(filled circles) to $n\p=1.45$, and then switched to increasing $\delta$
at constant $n\p$ as a potentially faster route to demixing. Indeed, at
$\delta\approx 8\%$ there was a smooth change in $p(n)$ from
single to double peaked; an example of the double peaked form is shown
in Fig.~\ref{fig:dendistsall}. The two associated phases were identified
as being fcc solids. As is physically reasonable, the
higher density solid (HDS) daughter  phase contains a surplus of the
smaller particles while the lower density solid (LDS) phase has more
of the larger particles.

\begin{figure}[t]
\includegraphics[type=pdf,ext=.pdf,read=.pdf,width=0.8\columnwidth,clip=true]{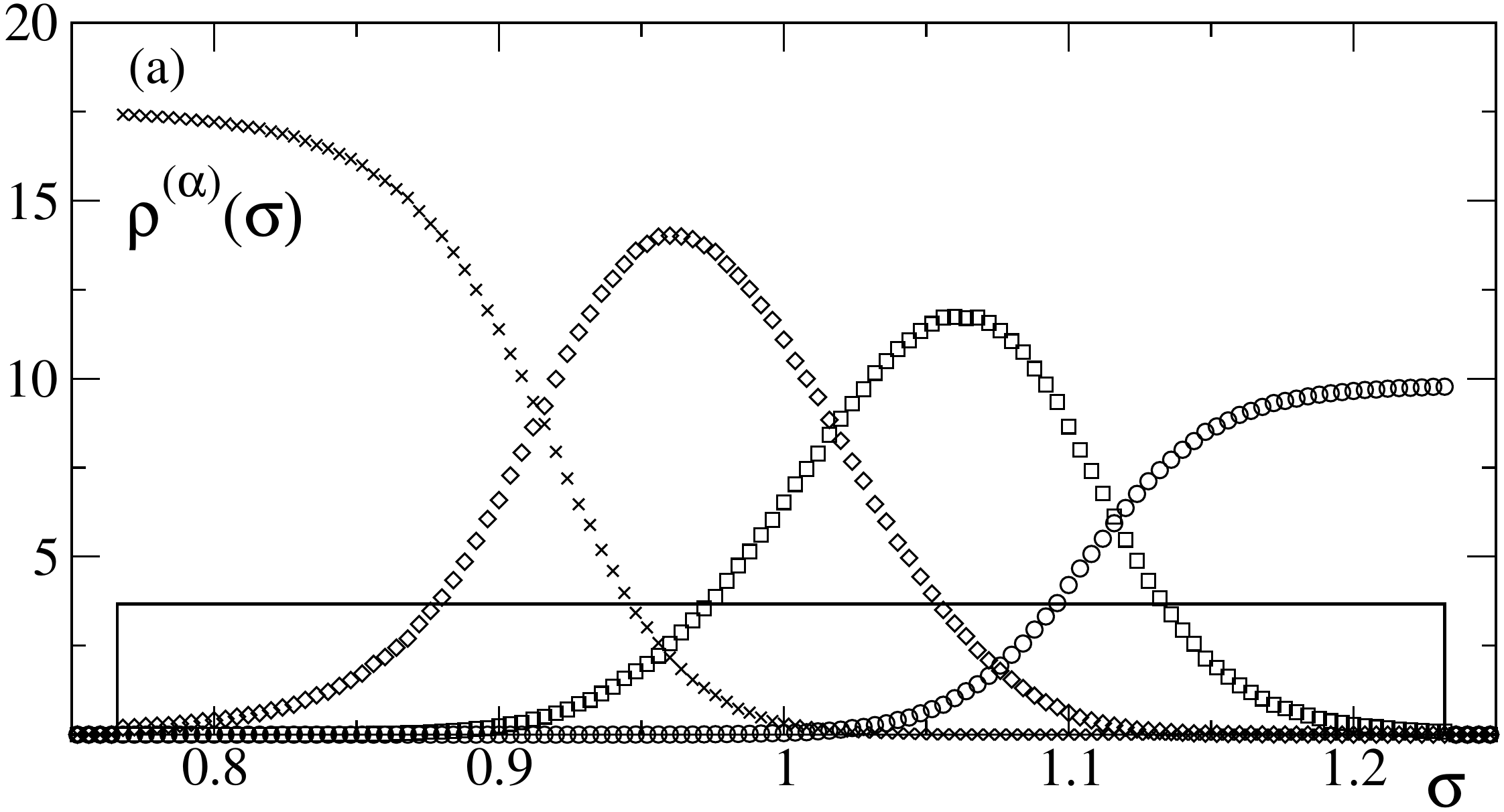}\\
\includegraphics[type=pdf,ext=.pdf,read=.pdf,width=0.8\columnwidth,clip=true]{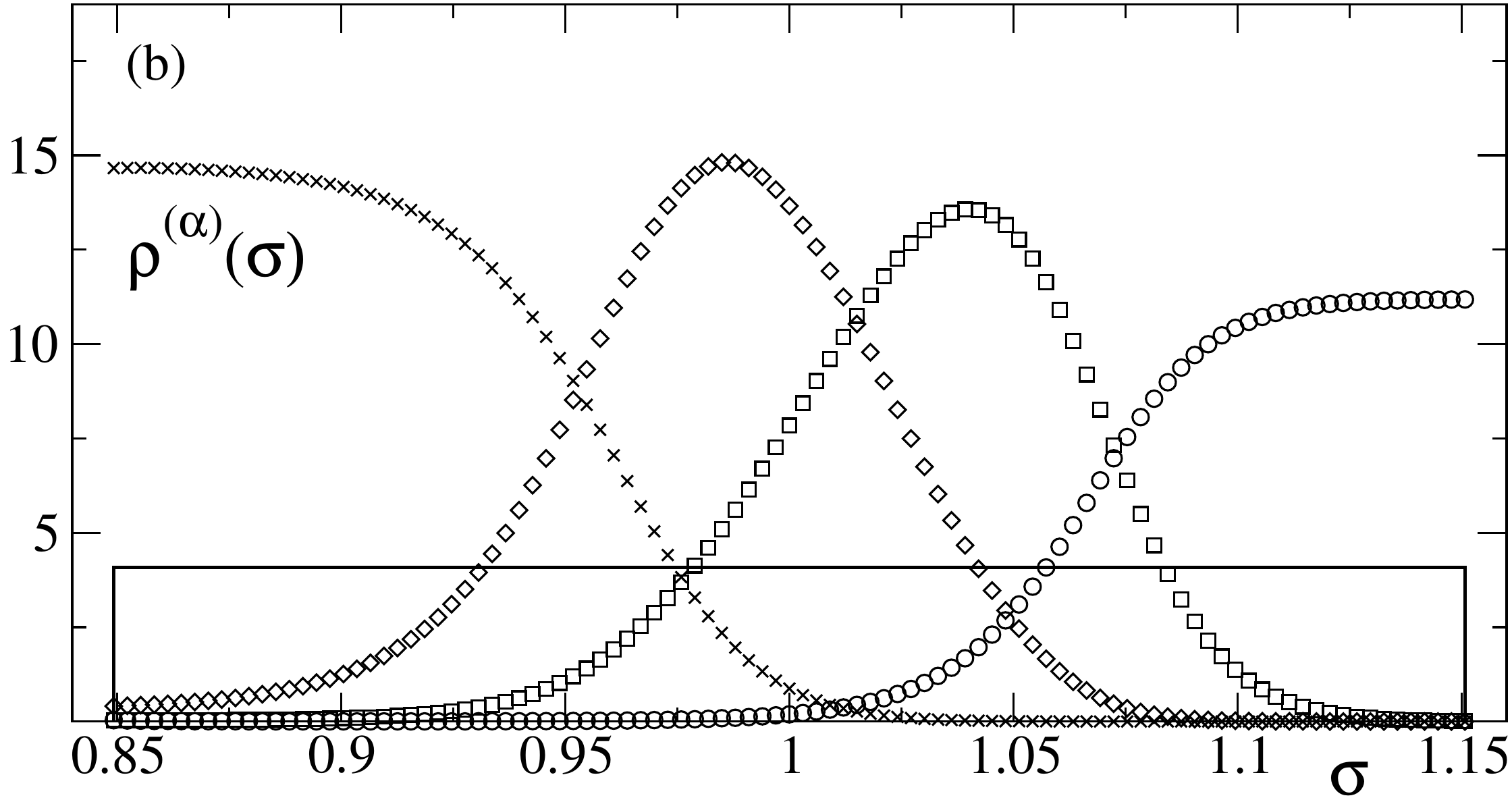}\\
\caption{{\bf (a)} Solid: parent distribution at ($n\p=1.73,
  \delta=13.5\%$). Symbols: Simulation results for the four daughter
distributions. The associated fractional
volumes $\lambda_\alpha$ are (left to right) 0.209, 0.188, 0.232,
0.373. 
% PKS weakened ``corresponding'' to ``comparable'' because we're not
% claiming anything as strong as corresponding states in liquid state theory
{\bf (b)} MFE results at the comparable state point ($n\p=1.232$,
$\delta=8.7\%$); fractional volumes are 0.273, 0.162, 0.200, 0.365.}

\label{fig:daughters4}
\end{figure}

Continuing to higher $\delta$ eventually led to spontaneous melting of
the system at $\delta=13.7\%$,  implying that the limit of metastability
with respect to a fluid-solid-solid (FSS) coexistence had been
overstepped, as is indeed predicted by our MFE calculations (see below).
 We therefore backtracked  slightly into the solid-solid (SS) region,
embarking on a new trajectory with increasing $n\p$ at constant
$\delta=13.5\%$. This produced a third peak in $p(n)$ at $n\p\approx
1.475$. The corresponding intermediate density solid (IDS) was again
found to be isostructural with the other two, with dominant particles
sizes between those in the HDS and LDS. Finally, increasing the overall
density to $n\p\approx 1.68$ we observed that the central IDS peak in
$p(n)$ split rather smoothly into two peaks, yielding a four peaked
structure (Fig.~\ref{fig:dendistsall}): four fcc solids now divide the
range of particles sizes among themselves (Fig.~\ref{fig:daughters4}).

We next compare to our theoretical MFE calculations. These used the same
parent size distribution (\ref{eq:th}) but, since no suitable
polydisperse model free energies are available for the soft repulsive
potential~(\ref{eq:softspheres}), the analysis was performed for hard 
spheres, using the methodology described elsewhere~\cite{Fasolo2004}. The qualitative
physics should be the same. Indeed, taking a  comparable path~\cite{path} through
the calculated phase diagram (Fig.~\ref{fig:part_pd}b) shows the same
features as in the simulations. (Quantitatively, the fluid-solid
coexistence region is narrower, and transitions to multiple solids occur
at lower $n\p$ and $\delta$, presumably because with a hard repulsion, a
crystal can accommodate above average-sized particles less easily.) Also
the fractionation effects are well reproduced, as shown in
Fig.~\ref{fig:daughters4}b for an SSSS state point at a location
comparable (relative to phase boundaries~\cite{path}) to the one in
Fig.~\ref{fig:daughters4}a.

\begin{figure}[h]
\includegraphics[type=pdf,ext=.pdf,read=.pdf,width=0.82\columnwidth,clip=true]{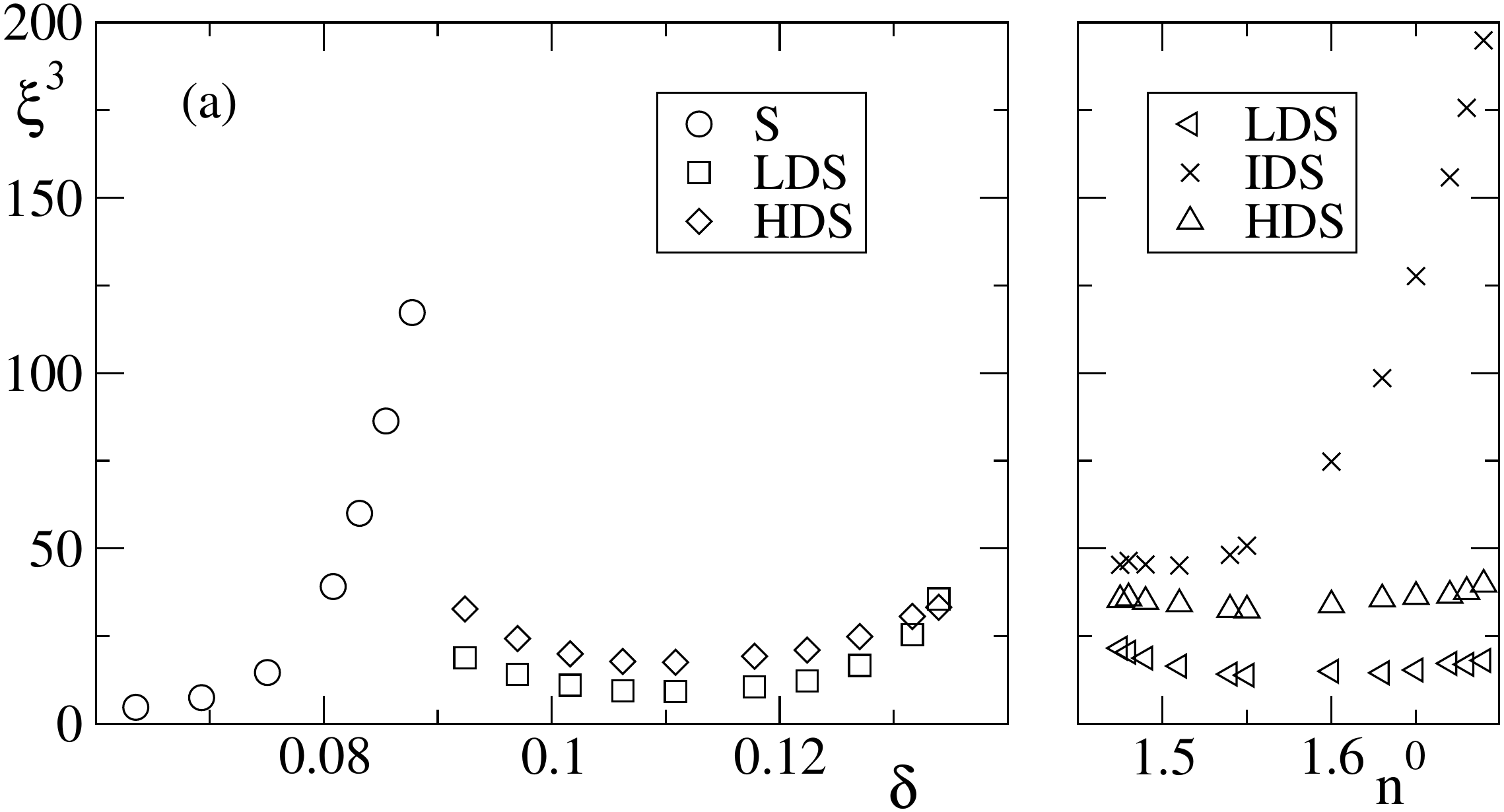}
\includegraphics[type=pdf,ext=.pdf,read=.pdf,width=0.82\columnwidth,clip=true]{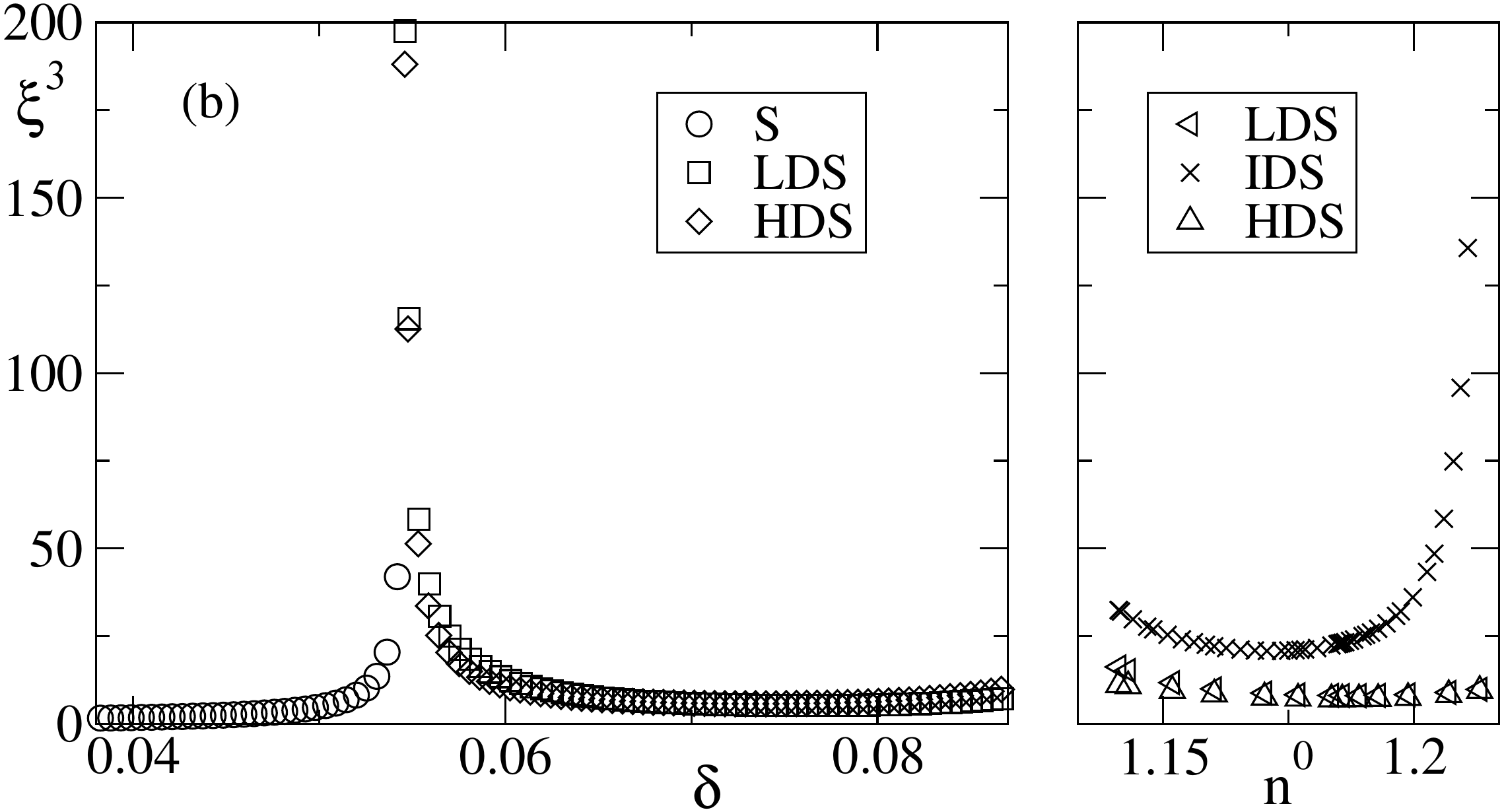}
  \caption{Correlation volume $\xi^3$ in the solid phases encountered along
  the phase diagram trajectories of Fig.~\ref{fig:part_pd}. {\bf (a)}
  Simulations, {\bf (b)} MFE calculations.}
\label{fig:fluctuation}
\end{figure}

%REVISION I took your footnote and incorporated it in the main text and
%tried to make it sound more accessible.....I also emphasised the
%differences one expects in the behaviour of the fluctuation on
%approaching 1st and 2nd order transitions.

A surprising feature of our results is that, from the variation of
$p(n)$, the transitions S$\to$SS and SSS$\to$SSSS appear to be
near-continuous in character, while SS$\to$SSS is strongly first order
as is usually expected for transitions in the solid state. A
near-continuous transition should be accompanied by size fluctuations
correlated over large distances, as precursors of the new phases,
whereas the fluctuations will remain small on approaching a first
order transition. To quantify these fluctuations we measure a
correlation volume $\xi^3$ from the variance across configurations  of
the mean particle size $\bar\sigma$. Suitably normalized, this variance, 
$\langle (\Delta\bar\sigma)^2\rangle$, is proportional to the spatial
integral over the pair correlation $g_{\sigma\sigma'}(\rv)$, weighted by
deviations of the particle sizes $\sigma$ and $\sigma'$ from the mean,
i.e.\ the correlation volume. In theoretical calculations, $\langle
(\Delta\bar\sigma)^2\rangle$ can be extracted from second derivatives of the
MFE~\protect\cite{Sollich2001}. Measurements of $\xi^3$ along the
trajectories through the phase diagrams are shown in
Fig.~\ref{fig:fluctuation}. This grows large near the transitions to two
and four solids, confirming their near-continuous character. In the
latter case, the splitting of the middle peak seen earlier in $p(n)$
suggests that the new solids arise out of the IDS phase, and this is
consistent with large fluctuations occurring (see
Fig.~\ref{fig:fluctuation}) only in this phase and not the HDS or LDS.
The MFE predictions are, again, in good qualitative accord with the
simulation data.

%REVISION. Added comment to the fact that similar findings observed for
%other parents within MFE

Our tailored simulations have provided a clear answer to long standing
questions surrounding the effect of size polydispersity on the
equilibrium phase behaviour of spherical particles: as density and/or
polydispersity are increased within the crystalline region, the system
demixes into an ever increasing number of fractionated fcc phases. Given
the high level of qualitative accord with MFE calculations, we are
confident that this scenario represents the true equilibrium situation. 
Since the MFE results are insensitive to whether the parent   
distribution is top hat (present work) or has a Schultz or triangular
form \cite{Fasolo2004}, we believe the demixing scenario to be quite 
general.
Understanding in detail when and why the demixing transitions are
near-continuous is an exciting open challenge.
Finally, we note that in spherical colloids demixing transitions
may not always be directly observable because fractionation requires
particle diffusion which is inhibited in solids. Nonetheless,
one might expect to see evidence for solid
demixing in regions where the solids coexist with a fluid (cf.\
Fig.~\ref{fig:part_pd}b) that can transport particles
to their preferred solid phase. Even in situations where
equilibrium cannot be reached for kinetic
reasons, knowledge of the true equilibrium state provides an important
baseline for interpreting dynamical effects \cite{Evans2001,PUSEY09}.

%\acknowledgments 
%Computational results were produced on a machine funded by HEFCE’s
%Strategic Research Infrastructure fund.

%TC:break biblio

\bibliographystyle{prsty}

\begin{thebibliography}{10}
\bibitem{path}
As standard mappings from soft to hard spheres fail at our densities, we
use phase diagram topology to identify comparable points. E.g.\ the
simulations at $\delta=13.7\%$ show an instability towards FSS very
close to the SS$\to$SSS transition. From the density difference between
these two points, relative to the separation between SS$\to$SSS and
SSS$\to$SSSS, we estimate the corresponding polydispersity for hard
spheres as $\delta=8.7\%$, just below the meeting point of the
SS$\to$FSS and SS$\to$SSS lines.
\end{thebibliography}
%TC:break supplementary

\end{document}